\documentstyle[twoside,fleqn,espcrc2]{article}

\newcommand{\AmS}{{\protect\the\textfont2
  A\kern-.1667em\lower.5ex\hbox{M}\kern-.125emS}}
\title{Monte Carlo Simulations of the SU(2) Vacuum Structure}

\author{ A.R. Levi
\address{
Center for Theoretical Physics, Laboratory for Nuclear Science
and Department of Physics, \\
Massachusetts Institute of Technology,
Cambridge, Massachusetts 02139 U.S.A.\\
{\it and} \ \ Department of Physics, Boston University, \\
590 Commonwealth Ave., Boston, Massachusetts 02215 U.S.A.}%
\thanks{
Work supported in part by the U.S. Department of Energy (DOE)
under contracts DE-AC02-76ER03069 and DE-FG02-91ER40676 }}

\begin{document}

\begin{abstract}
Lattice Monte Carlo simulations are performed for the SU(2)
Yang Mills gauge theory in the presence of an Abelian background
with external sources to obtain information on the effective potential.
The goal is to investigate the lowest Landau mode that, in
the continuum one-loop effective potential,
is the crucial mode for instability.
It is shown that also in the lattice formulation
this lowest Landau mode plays a very peculiar role, and it
is important for the understanding of the vacuum properties.
\end{abstract}

\maketitle

\section{INTRODUCTION}

To understand the IR behavior of non-Abelian gauge theory a
non-perturbative framework is necessary.
Therefore, lattice formulation is particularly useful.
However, the comparison between one-loop perturbative expansion
and lattice regularization might give important
information.
{}From this comparison we can learn about the
trustworthiness of perturbative expansion.
Furthermore, the loop expansion can provide indicative information
for the lattice quantities.

Despite the importance of Yang-Mills theories, the complete
solution of non-Abelian gauge theories
has yet to be found.
In order to gain a better understanding of these theories, a
necessary first step is the study of their vacuum structures.
Nevertheless, the vacuum structure of such theories has yet to be
understood  even in the simplest case, i.e. SU(2) without matter.
Moreover, the infrared properties of such theories must be
studied systematically if we want to have
some clue on confinement.

Many authors have studied one-loop effective potential for SU(2)
and the possible consequences \cite{savvidy,levi} .
In addition, several Monte Carlo simulations have already been done with a
wild range of techniques in 3 and 4 Dimensions\cite{levi,ambio}.

\section {ONE-LOOP APPROACH}

A powerful method to investigate the properties of Yang-Mills theories is
to compute of the effective potential
in the background gauge.
This method has been discussed extensively in the literature
\cite{dewitt}.
The technique is to
split the gauge field into a background ${\cal A}_\mu^b$
and a quantized field $\eta_\mu^b$, as
\begin{equation}
        A_\mu^b={\cal A}^b_\mu+\eta^b_\mu    ~,
\end{equation}
and subsequently performing a loop expansion.
This manifestly gauge invariant scheme is based on the observation that
the loop expansion corresponds to
an expansion in the parameter $\hbar$ which multiplies
the entire action. Hence, a shift of the fields or
a redefinition of the division of the Lagrangian
into free and interacting parts can be performed
at any finite order of the loop expansion without violating the
gauge invariance.
Nevertheless, there are several
subtleties on the exponentiation of the gauge constraint and in the
ghosts contributions to the finite part
of the effective action \cite{suzhou}.

We want to investigate the situation where ${\cal A}$ generates
a static constant chromomagnetic field.
A possible choice for ${\cal A}_\mu^b$
is the so called Abelian background:
\begin{equation}
     {\cal A}_\mu^b={1\over 2} H \delta^{b3}
     (x\delta_{\mu 2}-y \delta_{\mu 1})  ~.
\end{equation}
With this choice, the well known Savvidy
result for the one-loop effective potential is obtained
\begin{equation}
     E(H)={H^2\over 2}+{11 g^2H^2\over 48\pi^2} \biggl(
     \ln {gH\over \mu^2} - {1\over 2} - {i 6\pi \over 11}
     \biggr) +...
\end{equation}
The remarkable feature of this expression is that it
exhibits a minimum for $H$ different from zero.
However, as Nielsen and Olesen realized, due to the
imaginary part of the effective potential,
this minimum has unstable modes.
Since the above mentioned preliminary studies were done,
the property of this
non-trivial vacuum has been thoroughly investigated.
In particular a scenario, the so-called
``Copenhagen vacuum'' \cite{savvidy} was proposed.

However, in a strong
field configuration a perturbative analysis is unreliable,
and unstable configurations can only be analyzed by non-perturbative
methods. Therefore, the only
possible technique presently available to tackle this problem
is the one based on lattice regularization.

In this context it is interesting to note that the one-loop
effective potential can be obtained using
\begin{equation}
    V(H)=V_{classic}+const \sum \sqrt{\nu~}
\label{aaaa}
\end{equation}
\noindent
where $\nu$ are the eigenvalues of the second derivative of
the action with respect to the gauge fields, and
the summation is over all the eigenvalues.
This expression is a natural consequence of the saddle point
approximation:
\begin{equation}
    V(H)=V_{classic}-
          {\hbar \over 2 \Omega } \log\det
          \biggl({\delta^2 S \over \delta\eta\delta\eta}
          \biggr)+O(\hbar^2)
\end{equation}
where $\Omega$ is the volume factor.
The eigenvalue of
this problem can be found by realizing that there is
the same symmetry
as the one of the Landau levels problem, therefore
we have exactly the same class of solutions.
That yield to the eigenvalues
\begin{equation}
     \nu= k^2+(2 n + 1 + 2 S_z)gH
\end{equation}
\noindent
where $k=k_0^2 +k_z^2$, $n=0,1,2,...$ and $S_z=\pm 1$ is
the gluon polarization.

For the lowest landau mode, i.e. $n=0, S_z=-1$, we have $\nu<0$ that give
the imaginary part of the effective potential.
Note that with Abelian background it is possible to obtain
all eigenvalues positive only by insertion of non-gauge invariant terms
in the action.

\section{LATTICE RESULTS}

Until now we have argued that this lowest Landau mode plays
a very particular role on the one loop perturbative expansion.
This motivated us to understand what happened to this mode on
lattice where the task is non-perturbative.

To extract information on this mode on lattice we noted
that, due to the finiteness of the lattice $k_z$, is quantized as well,
with $k_z=2m \pi /L_z$, where $m$ is an integer.
Therefore, the lowest inhomogeneous $z$-mode, $m=1$,
becomes stable for lattices the extent of which
in the $z$-direction is smaller than
\begin{equation}
     L_z^{critic}={2\pi \over \sqrt{gH}} ~.
\end{equation}
This enables us to search for the critical size $L_z^{critic}$.
The homogeneous mode, $m=0$, which is always unstable, is eliminated by
imposing a delta condition
in the path integral.

We perform
Monte Carlo simulations that generate a background field
$H$ in the $z$ direction in the presence of
an Abelian source of strength $j$.

We use a heat bath updating procedure
with periodic boundary conditions and for the computational technique we
address to reference \cite{levi}.
To eliminate the homogeneous mode $m=0$
we force the Polyakov line in the $z$ direction to take a fixed value
different from zero.
Monte Carlo simulations are made with a lattice volume
$L^3*L_z$, where $L$ is the size of the $x,y,t$ directions.
The finite effect due to $L$ is not so crucial.
In fact, for the observable
that we are interested, it is sufficient to use $L=12-16$.
Simulations are made by varying $L_z$ from 4 to 50.

The expectation value of the
plaquette in the 1-2 plane $P[F_{12}]$ as a
function of $\beta$ and $j$ is measured.
We monitor the quantity
\begin{equation}
     X={P[F_{12}(\beta,0)]-P[F_{12}(\beta,j)]
      \over j P[F_{12}(\beta,0)}   ~,
\end{equation}
which is proportional to the main contribution of the vacuum energy
of the plaquette in the $z$ direction \cite{levi,ambio},
and is very sensible to the presence of the unstable mode.
The total vacuum energy obviously has also contributions from the other
oriented plaquette; however, in the region we are interested,
these other contributions are smoothly variable functions, and
thus they will not affect the critical behavior.

We evaluated $X$ for different values of $\beta$
and $j$ performing, for large $j$ 4500 sweeps after discharging 500 for
thermalization. For smaller $j$ we increase the number of
sweeps until a significant amount of data was collected.

Our data \cite{levi} show that there
is no sign of the unstable mode away from
the critical $\beta$ region ($\beta=2.1 - 2.5$).
The situation changes dramatically in the critical region where
the instability appears as a decrease of the vacuum
energy contribution to the plaquette in the $z$ direction.
This effect becomes more evident in the presence of
strong sources.
The fact that the instability disappears outside of the scaling window
is a strong evidence that the
instability is a distinguishing feature of the continuum
rather than the strong coupling vacuum.
In order to study the IR property of the continuum
theory we should remain in the region where the instability is manifest.
The disappearance of the instability as $\beta\to0$ may give a clue to
understanding the difference between the physics of the continuum and the
strong coupling lattice theory.

We systematically analyzed the critical region of $\beta$
for several values of $j$.
The values of $L_z^{critic}(\beta,j)$ are obtained
by interpolating the kink of $X$ and taking the median value.
$L_z^{critic}$ was found to be dependent on the source $j$
and on $\beta$ in this region according to the renormalization group
dependence.
It is clear from our data that $L_z^{critic}$
belongs to the confinement phase and that
there is good agreement with the renormalization
group equation.
Our data follow the same dependence of the deconfinament
transition \cite{kuti} as should be for a dimensional scale length.
Hence the ratio between
the $L^{critic}$ and the deconfinament scale
parameter $L_{dec}$ is independent of $\beta$.
The relevant physical quantities must be obtained in the limit
of vanishing of the induced field $Q(j) \rightarrow 0$.
{}From our data we obtain
\begin{equation}
       {L_z^{critic}\over L_{dec}} = 2.5 \pm 0.2.
\end{equation}
Using the known value of $L_{dec}$  \cite{kuti} we can establish
that the lattice system should be greater than $L_z \sim 18$  in order to
reflect the richness of the non-perturbative vacuum of the continuum theory.

In addition to this new scale there is a sequence of
$m$ modes. In fact for each $m$ we have:
\begin{equation}
     {2\pi m \over L_{z(m)}^{critic}}=\sqrt{gH}
     ~~~~\Rightarrow ~~~~ L_{z(m)}^{critic}
       \sim {2.5 \over T_{dec}} m ~.
\label{bbbb}
\end{equation}
The simulations show that these modes follow with
good approximation eq.~(\ref{bbbb}). This is strong evidence for the
harmonicity of this modes.
This is a quite surprising result because we are in a region
of strong non-perturbative effects.
It is stimulating to think in terms of a parallel between the
situation described above
and the integer Quantum Hall Effect.
This similarity is based on the dual picture
($ z \leftrightarrow t $, $ E \leftrightarrow H $, etc...),
and on the plateau structures for each mode.

\noindent
{\bf Acknowledgements}

It is a pleasure to thank
Janos Polonyi and Suzhou Huang
for several useful discussions.

\end{document}